\documentclass[11pt]{article}
\usepackage{amsmath}
\usepackage{graphicx} 
\usepackage{epsf}
\usepackage{graphics}
\usepackage[latin1]{inputenc}
\newcommand{\be}{\begin{equation}}
\newcommand{\ee}{\end{equation}}
\newcommand{\bea}{\begin{eqnarray}}
\newcommand{\eea}{\end{eqnarray}}

\title{The long wavelength  limit of hard thermal loop effective actions}
\author{F. T. Brandt$^1$, J. Frenkel$^1$ and J. C. Taylor$^2$
\\ \\ $^1$ Instituto de F\'{\i}sica, Universidade de S\~ao Paulo,\\ 
Rua do Mat\~ao, Travessa R, 187, S\~ao Paulo, SP 05508-090, Brazil 
\\ \\ $^2$ DAMTP, Centre for Mathematical Sciences,\\ $\,$ Wilberforce Road, Cambridge, CB3 0WA, United Kingdom}

\begin{document}

\maketitle

\subsection*{Abstract}
We derive a closed form expression for the long wavelength limit of the effective action
for hard thermal loops in an external gravitational field. It is a function of the metric, independent
of time derivatives. It is compared and contrasted with the static limit, and with the
corresponding limits in an external Yang-Mills field.

\section{Introduction}

In perturbative thermal field theory, in order to partly control infra-red divergences, it is necessary
first to calculate hard thermal loops, where external momenta are much less than the
temperature $T$ \cite{Braaten:1990mz,Frenkel:1990br}. 
The effective actions for the hard thermal loops enjoy some nice
properties \cite{Brandt:1993mj}. They are non-local, gauge-invariant (not just BRST invariant) functionals
of external Yang-Mills or gravitational fields.

There are two special cases: the static limit (where the external fields are time-independent)
and long wave-length  (LWL) limit (where the external fields are position independent).
In both these limits, the effective actions turn out to be functions (not functionals) of the
external fields, with no derivatives at all. That is, in momentum space, the static limit
is independent of momenta as well as energies, and the LWL limit is independent of energies
as well as momenta. Nevertheless, these two limits give different functions.

\section{External Yang Mills field}
For hard thermal loops in an external Yang Mills field, both the static and the LWL limits
turn out to be non-zero for the 2-point function only. This may be
written, in general, in the form 
\cite{lebellac:book96}
\be \label{eq1}
\frac{{\cal C}_{\mbox{{\scriptsize YM}}}\,g^2 T^2}{12\pi}\int{\rm d}\Omega
\left[k_0\, \frac{Q_{\mu}Q_{\nu}}{k\cdot Q} - n_{\mu}n_{\nu}\right]
\ee
where ${\cal C}_{\mbox{{\scriptsize YM}}}$ is a Casimir for the internal particles, $g$ is the coupling-constant, 
$k$ is the external field momentum, $n=(1;0,0,0)$,
$Q=(1;{\bf Q})$ with $|{\bf Q}|=1$,  and the integration is over its directions.
 
The static limit ($k_0=0$) is obtained immediately from \eqref{eq1}. 
It comes from the second term alone, and is clearly independent of
${\bf k}$. It is easy to verify that this result is invariant
under gauge transformations independent of time.
 
In the LWL limit, $k_0/k\cdot Q=1$, and the angular integration is trivial and the two terms in \eqref{eq1}
cancel unless both $\mu$ and $\nu$ are space-like. The contribution is then
independent of $k_0$, and differs by a factor $-1/3$ from the static limit.
It is again easy to check that the result is gauge invariant under gauge transformations
independent of position. This suggests that it is the complete effective action in the
LWL limit.

One can in fact show that all higher point functions are zero in either the static or LWL limits.
For the static case, this follows from equation (8) of reference \cite{Taylor:1990ia}.
  
For the LWL limit, we can proceed as follows. Take first the 3-point function.
The Ward identities (which are known to apply to hard thermal amplitudes \cite{Frenkel:1990br})
state
\be\label{eq2}
k_0\,\Gamma_{0,\mu ,\nu}(k,p,q)=g \Gamma_{\mu,\nu}(p)-g\Gamma_{\mu,\nu}(q)
\ee
Since the two terms on the right are independent of momenta in the LWL
limit, all components of
the 3-point function vanish except perhaps the purely spatial indices. Further, the function 
involves angular integral like (see for example \cite{Frenkel:1990br}) (here $a,b,c$ are colour indices)
\be\label{eq3}
\int{\rm d}\Omega F^{abc}(k,p,q,k\cdot Q, p\cdot Q,q\cdot Q)Q_{\lambda}Q_{\mu}Q_{\nu}
\ee
In the LWL limit, the function $F$ is independent of ${\bf Q}$ and therefore of $Q$. So the $0,0,0$
component (for example) of the above tensor is non-zero unless the function $F=0$
itself. It follows that all components of the 3-point function vanish
in the LWL limit.
  
Actually, the part of the 3-point function with all indices spatial must vanish in the LWL
limit anyway, by rotational invariance. However, we give the argument in the above form
because it generalizes in an obvious way to the 4-point function and higher.
  
Up to the 4-point function, the LWL limit can also be found explicitly
from the expressions given in reference \cite{Frenkel:1990br}.
%  
%  [Mention Boltzmann?]
  
\section{External gravitational field}
  
For hard thermal loops in an external gravitational field, 
the static limit is well known \cite{Rebhan:1990yr}
and can be written in terms of an effective action as
\be\label{eq4}
S_{\mbox{\scriptsize{S}}}=\frac{{\cal C}\,\pi^2\,T^4}{90} 
\int {\rm d}^4 x 
\sqrt{-g} (g_{00})^{-2},
\ee
where the Casimir ${\cal C}$ gives the number of internal degrees of freedom
(for internal bosons; for internal fermions there is an extra factor of $1/2$)
and the time integration is taken over a finite range, 
since the integrand is time-independent.
 % [???? Casimir???]
 This is invariant under time-independent gauge transformations, and under Weyl transformations.

We have not been able to find a similar explicit closed form for the
LWL limit. But there is
an implicit closed form representation, inspired by a Boltzmann equation approach
\cite{Brandt:1995mv}. 
We will first state this form, verify that it is invariant under gauge (that is
coordinate transformations) and Weyl transformations, and finally prove that it is the
unique, correct result.

The form of the effective action in this limit is
\bea\label{eq5}
S_{\mbox{\scriptsize{LWL}}} &=& \frac{{\cal C}}{(2\pi)^3}
\int {\rm d}^4 x 
\int {\rm d}^4 p\, \theta(p_0)
\left[\theta(g^{\mu\nu}p_{\mu}p_{\nu})-\theta(\eta^{\mu\nu}p_{\mu}p_{\nu})\right]
N\left(\sqrt{p_ip_i}/T\right)
\nonumber \\
&=&-\frac{{\cal C}\,\pi T^4}{120} 
\int {\rm d}^4 x \int{\rm d}\Omega 
\left[\sqrt{\left(\frac{Q_i g^{0i}}{g^{00}}\right)^2-\frac{g^{ij}Q_iQ_j}{g^{00}}}-1\right]
\eea
where $N$ is the Bose distribution function 
(if the internal particles are fermions, an extra factor $1/2$ 
should be inserted in the second line of \eqref{eq5}) and
the space integral is taken to be over a finite volume, since the 
integrand is space-independent. This form is suggested because
$p_i$ is a solution of the Boltzmann equation ((3) of \cite{Brandt:1995mv} ) in the
LWL limit. Also, $p_i$
is invariant under space-independent gauge transformations.
The above function has three properties. It is gauge invariant under gauge
transformations independent of position and it it is Weyl invariant. Furthermore, if we
expand in powers of $h$ where
\be\label{eq6}
\sqrt{-g}g^{\mu\nu}=\eta^{\mu\nu}+h^{\mu\nu}
\ee
the coefficients of
\[
h^{\mu\nu}h^{\lambda\sigma} \cdots
\]
are totally symmetric in $\mu,\nu,\lambda,\sigma,\cdots$.
We will now show, from the analysis of the hard thermal perturbation
theory, that these three
properties are required, and that they suffice to determine the HTL uniquely.
It thus follows that \eqref{eq5} is correct.

We shall take as an example the 3-graviton term in perturbation theory, but try
to express the argument so it is clear how it generalizes to higher orders.
According to \cite{Brandt:1992qn}, the 3-graviton amplitude involves a angular integration (as in \eqref{eq5})
of a tensor
\be\label{eq7}
C_{\mu\lambda,\alpha\beta,\rho\sigma}(Q,k, k',k'').
\ee
The tensor could be proportional to some product of the vectors 
$Q_{\mu},\, k_{\mu},\,k'_{\mu},\,k''_{\mu}$ and the Minkowski metric $\eta_{\mu\nu}$.

The tensor $C$ in \eqref{eq7} obeys a Ward identity connecting it to the 2-graviton tensor:
\bea\label{eq8}
2 k^\mu\, C_{\mu\lambda,\alpha\beta,\rho\sigma}(Q,k, k',k'')&=&
\left[-k_\alpha\, C_{\lambda\beta,\rho\sigma}(Q,k'')-\left(\alpha\leftrightarrow\beta\right)+
\right. \nonumber \\ & & \left.
k'_\lambda\, C_{\alpha\beta,\rho\sigma}(Q,k'')\right]
+\left[k',\alpha,\beta \leftrightarrow k'',\rho,\sigma\right]
\nonumber \\ & & 
\eea
and also a Weyl identity
\be\label{eq9}
\eta^{\rho\sigma} C_{\mu\lambda,\alpha\beta,\rho\sigma}(Q,k, k',k'')=
-C_{\mu \lambda,\alpha \beta}(Q,k)-C_{\mu \lambda,\alpha \beta}(Q,k')
\ee
These identities are in fact sufficient to determine C in \eqref{eq7}. But for the moment we deduce
only that the tensor $C$ in \eqref{eq7} does not have any terms containing $\eta_{\mu\nu}$.
The reason is, firstly, that the 2-graviton term on the right of the Ward identity \eqref{eq8}
do not contain $\eta_{\alpha\beta}$ etc., and secondly that there is no $k_{\lambda}$
on the right of \eqref{eq8} as there would be if \eqref{eq7} contained $\eta_{\mu\lambda}$.
This argument can be continued iteratively to all orders.

We now specialize to the LWL limit. Then the tensor $C$ in \eqref{eq7} must be proportional
to some product of the vectors $Q$ and $n$ (and not the tensor $\eta$).
Also, the Ward identity \eqref{eq8} becomes
\be\label{eq10}
2\,C_{0\lambda,\alpha\beta,\rho\sigma}=-[
n_{\alpha} C_{\lambda\beta,\rho\sigma}+
(\alpha \leftrightarrow \beta)+
\frac 1 2 n_{\lambda} C_{\alpha\beta,\rho\sigma}
]-(\alpha,\beta\leftrightarrow\rho,\sigma)
\ee
Thus all components of the 3-point function are determined, and are independent of energies,
except the one with purely spatial indices.
 
Next, Weyl invariance \eqref{eq9} determines the trace $C^{\sigma}_{\sigma\mu\lambda,\alpha\beta}$, and therefore
the spatial trace $C_{ii,kl,mn}$ is also determined. But, $C_{ij,kl,mn}$ is proportional to
$Q_iQ_jQ_kQ_lQ_mQ_n$ and so it is totally symmetric; so if its trace is determined
the complete tensor is determined.
 
%% The lagrangian in \eqref{eq5} is gauge and Weyl invariant.
These arguments can be generalized to all orders. We infer that the
LWL limit is uniquely determined
by three conditions: the Ward identity \eqref{eq8}, the Weyl
identity \eqref{eq9}, and the complete symmetry
of the purely spatial parts of the $C$ tensors (which is a consequence of the absence of
the $\eta$ tensor). The conjectured form \eqref{eq5} satisfies all these three conditions, so it must be
the correct unique LWL limit.
(It must satisfy the Ward identities \eqref{eq9} 
because the first line of \eqref{eq5} is explicitly gauge invariant.)
In fact, up to the 4-point function, we have explicitly
verified that \eqref{eq5} consistently generates, in the LWL limit,
the perturbative hard thermal loops in a gravitational field.

\bigskip
{\bf Acknowledgments}

\noindent
F. T. B. and J. F. would like to thank FAPESP and CNPq (Brasil) for a grant.

%\bibliographystyle{elsart-num}
%\bibliography{all_new}

\end{document}